\documentclass[epj]{svjour}
\usepackage{graphics}
\usepackage[dvipsnames]{xcolor}

\newcommand*\linkcolours{ForestGreen}

\usepackage{graphicx}
\usepackage{amssymb}
\usepackage{amsmath}
\usepackage[normalem]{ulem}
\usepackage{url,hyperref}
\usepackage{subfigure}
\usepackage[numbers,sort&compress]{natbib}

\hypersetup{
colorlinks,
linkcolor=\linkcolours,
citecolor=\linkcolours,
filecolor=\linkcolours,
urlcolor=\linkcolours}
\usepackage{float}
\usepackage{epstopdf}

\begin{document}
\title{Deuteron and Antideuteron Coalescence in Heavy-Ion Collisions: Energy Dependence of the Formation Geometry}
\author{Apiwit Kittiratpattana\inst{1}, Michael F. Wondrak\inst{2,3}, Medina Hamzic\inst{4}, Marcus Bleicher\inst{2,3}, Christoph Herold\inst{1}, Ayut Limphirat\inst{1}
}
\institute{School of Physics and Center of Excellence in High Energy Physics \& Astrophysics,  Suranaree University of Technology, Nakhon Ratchasima  30000, Thailand, \email{apiwitkitti@g.sut.ac.th, herold@g.sut.ac.th} \and 
Institut f\"ur Theoretische Physik, Johann Wolfgang Goethe-Universit\"at, Max-von-Laue-Str.~1, 60438 Frankfurt am Main, Germany \and 
Helmholtz Research Academy Hesse for FAIR, Campus Frankfurt, Max-von-Laue-Str.~12, 60438 Frankfurt am Main, Germany \and 
University of Sarajevo, Obala Kulina bana 7/II, 71000 Sarajevo, Bosnia and Herzegovina}
\titlerunning{(Anti-)Deuteron Coalescence: Energy Dependence of the Formation Geometry}
\authorrunning{A. Kittiratpattana \textit{et al.}}
\date{Received: date / Revised version: date}
%
\abstract{
We investigate the collision energy dependence of deuteron and antideuteron emission in the RHIC-BES low- to mid-energy range $\sqrt{s_{NN}} = 4.6-200$~GeV where the formation rate of antinuclei compared to nuclei is strongly suppressed. In the coalescence picture, this can be understood as bulk emission for nuclei in contrast to surface emission for antinuclei. By comparison with experimental data on the coalescence parameter $B_2$, we are able to extract the respective source geometries. This interpretation is further supported by results from the UrQMD transport model, and establishes the following picture: At low energies, nucleons freeze out 
over the total fireball volume, while antinucleons are annihilated inside the nucleon-rich fireball and can only freeze out on its surface. Towards higher energies, this annihilation effect becomes irrelevant (due to the decreasing baryochemical potential) and the system's freeze-out is driven by the mesons. Thus, the nucleon and antinucleon freeze-out distributions become similar with increasing energy. 
\keywords{deuteron formation via coalescence -- antinuclei suppression -- heavy-ion collisions}
\PACS{
   {25.75.-q}{Relativistic heavy-ion collisions}  \and
   {27.10.+h}{Deuterons}
   } 
} 
\maketitle

\section{Introduction}
Among the four fundamental forces in nature, the strong interaction, which is described by quantum chromodynamics (QCD), governs the physics on subatomic scales. Probing these is not only challenging experimentally but also difficult theoretically due to the non-abelian nature of QCD. Heavy-ion collisions provide a tool to probe nuclear matter under extreme temperatures and densities by colliding two nuclei in the accelerator. The fireball created in such a collision heats up to form a quark-gluon plasma (QGP), then expands and subsequently hadronizes into particles that are measured in the detector. 

The exploration of cluster formation in nuclear reactions started more than 50 years ago \cite{Hagedorn:1960zz, Butler:1961pr} and has been continued since then \cite{Nagle:1996vp,SchaffnerBielich:2000wj,Monreal:1999mv,Chen:2003ava,Oh:2007vf,Zhu:2015voa,Chen:2018tnh}. The physics of light (anti-)nuclei has already a long history covering a broad scientific range from astrophysics, \textit{e.g.}, Big Bang nucleosynthesis \cite{Malaney:1993ah}, hypermatter in neutron stars \cite{Rufa:1990zm} or signals of dark matter annihilation \cite{Carlson:2014ssa,Korsmeier:2017xzj} to nuclear-particle physics \cite{Braun-Munzinger:2018hat} and even physics beyond the standard model \cite{Beyer:1993zw, Aid:1995kf}. The formation process of nuclear clusters in heavy-ion collisions is however still debated. Essentially two ideas exist to describe the measured yields: direct thermal emission of the bound cluster from the chemical freeze-out surface or coalescence of the emitted baryons at kinetic freeze-out \cite{Mrowczynski:2019yrr,Braun-Munzinger:2018hat,Sombun:2018yqh,Bazak:2018hgl,Mrowczynski:2016xqm}. 

In this work, we explore antideuteron and deuteron formation by coalescence to reconstruct the spatial geometry of the emission source, considering that at low energies, the formation of antideuterons is strongly suppressed compared to deuterons. A successful ansatz to explain this discrepancy assumes different emission regions \cite{Mrowczynski:1993cx} for nucleon and antinucleon sources \cite{Nagle:1994wj,Nagle:1996vp} in contrast to simple coalescence models which do not consider spatial differences. The main assumption \cite{Bleicher:1995dw, Mrowczynski:1993cx} is that the antinucleons can only be emitted from the outer shell of the medium due to the huge nucleon-antinucleon annihilation cross section in the central baryon-rich region. Nucleons, on the other hand, are emitted over the whole volume. For this paper we will follow the specific implementation of this idea suggested by Mr\'{o}wczy\'{n}ski \cite{Mrowczynski:1993cx}, assuming spherically symmetric (anti-)nucleon source functions and call this approach "Mr\'{o}wczy\'{n}ski coalescence model".

In this paper, we use this model to systematically investigate the energy dependence of the size of the nucleon and antinucleon emission sources. We determine the different shapes of the antinucleon and nucleon emission regions as a function of collision energy. Our paper is structured as follows: 
In sect.~\ref{sec:model}, we explain the idea of Mr\'{o}wczy\'{n}ski's coalescence model in detail. This is followed by our results on the extracted source radii in sect.~\ref{sec:result1}. These are then compared to radii obtained from the UrQMD model in sect.~\ref{sec:urqmd}. Finally, we close with a summary and conclusions in sect.~\ref{sec:concl}.

\section{(Anti-)Deuteron Formation Rate and Source Geometry}
\label{sec:model}
The coalescence model describes the formation of baryonic clusters in the freeze-out stage of a heavy-ion collision. A pair of final-state (anti-)nucleons carrying similar momenta can coalesce to form a deuteron or an antideuteron with total momentum $\textbf{P}$. The invariant differential production cross sections for deuterons ($d$) and nucleons ($p$) -- and accordingly for antideuterons and antinucleons -- are related by
\begin{equation}
  E\frac{d^3\sigma_d}{dP^3} = \frac{B_2}{\sigma_{inel}} \left( \frac{E}{2}\frac{d^3\sigma_p}{d\left(P / 2\right)^3} \right)^2~,
\end{equation}
where $(E,\textbf{P})$ and $(E/2,\textbf{P}/2)$ denote the deuteron and nucleon 4-momenta and $\sigma_{inel}$ is the total inelastic cross section. The coalescence parameter $B_2$ can be measured in heavy-ion experiments or obtained from transport or coalescence models. 
Here, we employ the spatial coalescence approach defined by Mr\'{o}wczy\'{n}ski \cite{Mrowczynski:1993cx}. 
It is based on the formation rate, $\mathcal{A}\equiv \frac{m}{2} B_2$, with $m$ denoting the nucleon mass (neglecting the mass difference between protons and neutrons). The formation rate $\mathcal{A}$ is calculated as 
\begin{equation}
\label{eq:A}
  \mathcal{A} \equiv \frac{3}{4}(2\pi)^3 
  \int \int d^3 r_1 d^3 r_2
  \mathcal{D}(\textbf{r}_1) \mathcal{D}(\textbf{r}_2)
  |\psi_d(\textbf{r}_1, \textbf{r}_2)|^2~,
\end{equation}
where the nucleon source function $\mathcal{D}(\textbf{r}_i)$ describes the probability of finding a nucleon at a given point $\textbf{r}_i$ at kinetic freeze-out. Furthermore,  $\psi_d(\textbf{r}_1, \textbf{r}_2)$ denotes the deu\-ter\-on wave function. The nucleons are assumed to be emitted simultaneously and uncorrelated. 

We use the ansatz that the nucleon source is distributed over the whole fireball (volume emission), while the antinucleon source is suppressed towards the center of the fireball (surface emission). The nucleon source function $\mathcal{D}(\textbf{r})$ is parametrized by a normalized Gaussian \cite{Mrowczynski:1993cx},
\begin{equation}
\label{eq:D}
  \mathcal{D}(\textbf{r}_i) = \frac{\exp{\left( - \textbf{r}_i^2/2r_0^2 \right)}}{(2\pi)^{3/2} r_0^3}~,
\end{equation}
with $r_0$ being the radius of the fireball. The normalized antinucleon source function $\bar{\mathcal{D}}$ contains a second Gaussian of width $r_*$ that effectively cuts out the central region,
\begin{equation}
\label{eq:Dbar}
 \bar{\mathcal{D}}(\textbf{r}_i) =
   \frac{ \exp{\left( -\textbf{r}_i^2/2r_0^2 \right)}
    - \exp{\left( -\textbf{r}_i^2/2r_*^2 \right)} }{(2\pi)^{3/2} (r_0^3-r_*^3)}~.
\end{equation} 

\begin{figure}[!ht]
\centering
\includegraphics[width=\columnwidth]{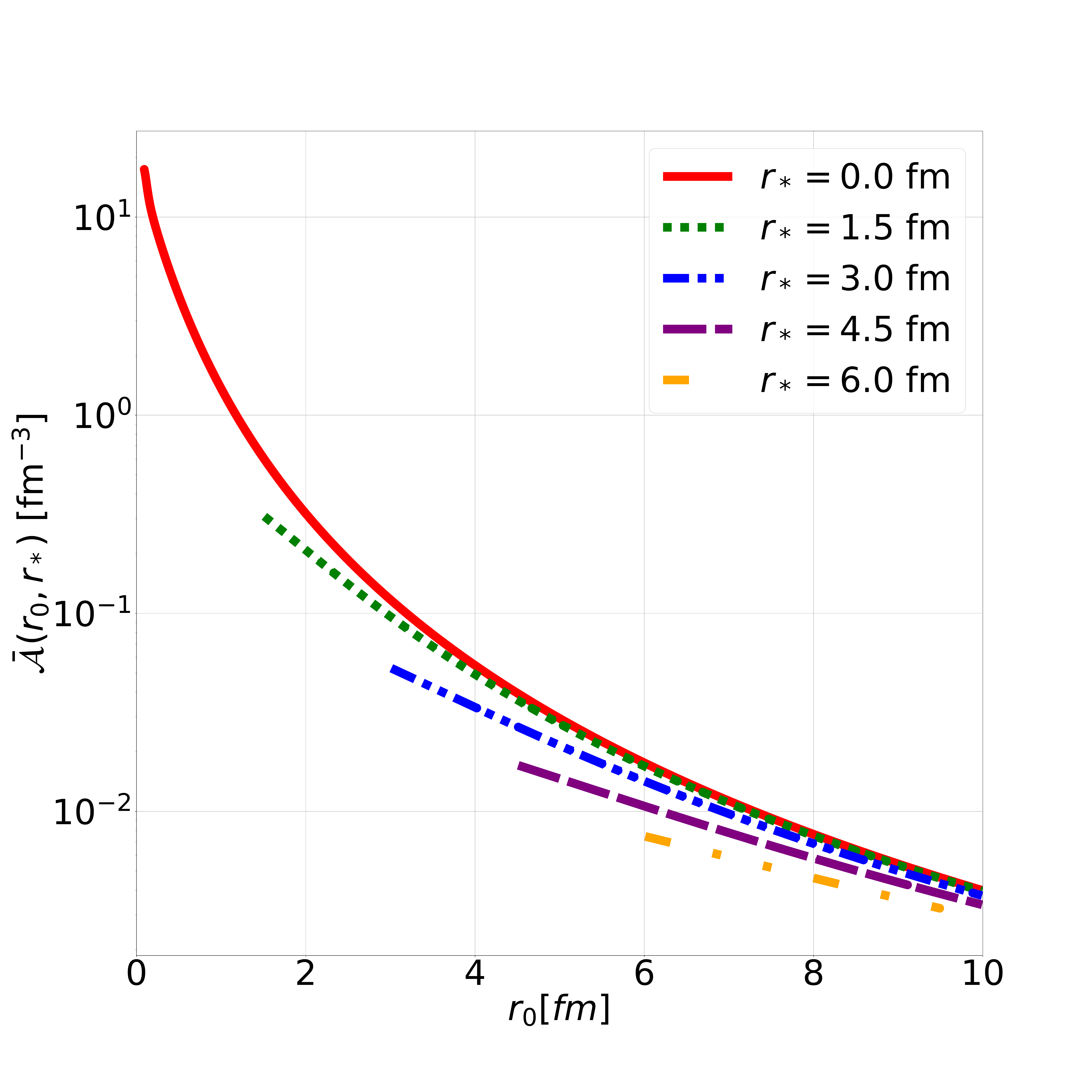}
\caption{(Color online) Antideuteron formation rate as a function of the source radius $r_0$ as obtained from the coalescence model for several values of the suppression radius $r_*$.}
\label{fig:formationrate}
\end{figure}

It is convenient to formulate the integral in eq.~\eqref{eq:A} in center-of-mass coordinates $\textbf{P}= \textbf{p}_1+\textbf{p}_2$, $\textbf{R}=\frac{1}{2}\left( \textbf{r}_1+\textbf{r}_2\right)$ and relative coordinates  $\textbf{p}=\frac{1}{2}\left(\textbf{p}_1-\textbf{p}_2\right)$, $\textbf{r}=\textbf{r}_1-\textbf{r}_2$. The deuteron wave function then factorizes to
\begin{equation}
  \psi_d(\textbf{r}_1, \textbf{r}_2) = \text{e}^{i\textbf{P}\cdot \textbf{R}}\phi_d(\textbf{r})~,
\end{equation}
with the Hulth\'en wave function,
\begin{equation}
\label{eq:wavefn}
  \phi_d(\textbf{r}) = \left( \frac{\alpha\beta(\alpha+\beta)}{2\pi(\alpha-\beta)^{2}}\right)^{1/2}\frac{\text{e}^{-\alpha r}-\text{e}^{-\beta r}}{r}~,
\end{equation}
where $\alpha=0.23$~fm$^{-1}$ and $\beta = 1.61$~fm$^{-1}$ \cite{hodgson1971nuclear}.
For the formation rate $\mathcal{A}$ in relative coordinates, we write
\begin{equation}
\label{eq:Ar}
  \mathcal{A} \equiv \frac{3}{4}(2\pi)^3 
  \int d^3 r \mathcal{D}_r(\textbf{r}) |\phi_d(\textbf{r})|^2~,
\end{equation}
with the nucleon relative source function
\begin{equation}
\label{eq:Dr}
  \mathcal{D}_r( \textbf{r} ) = \frac{1}{(4\pi)^{3/2}r_0^3}\exp{ \left( -\textbf{r}^2/4r_0^2 \right)}~,
\end{equation}
which only depends on the relative coordinate $\textbf{r}$. The antinucleon product source function\footnote{Please note that this form differs from the one given in \cite{Mrowczynski:1993cx} which contained a minor mistake.} in relative coordinates is then
\begin{equation}
\label{eq:Drbar}
 \bar{\mathcal{D}}_r(\textbf{r}) =
   \frac{ r_0^3 \hspace{1mm}\text{e}^{-\frac{\textbf{r}^2}{4r_0^2}} + r_*^3 \hspace{1mm}\text{e}^{-\frac{\textbf{r}^2}{4r_*^2}} - \frac{2^{\frac{5}{2}}r_0^3 r_*^3}{(r_0^2+r_*^2)^{\frac{3}{2}}}\hspace{1mm}\text{e}^{-\frac{\textbf{r}^2}{2\left(r_0^2+r_*^2\right)}} }{(4\pi)^{\frac{3}{2}} (r_0^3-r_*^3)^2}~.
\end{equation}
Figure \ref{fig:formationrate} shows the antideuteron formation rate $\bar{\mathcal{A}}(r_0, r_*)$ according to eq.~\eqref{eq:Ar} and \eqref{eq:Drbar} for different values of $r_0$ and $r_*$. Note that for $r_*=0$ (disappearance of the annihilation region) the formation rate of antideuterons is the same as for deuterons.

\section{Results}
\label{sec:result1}
\begin{figure}[!ht]
\centering
  \begin{subfigure}{}
    \includegraphics[width=\columnwidth]{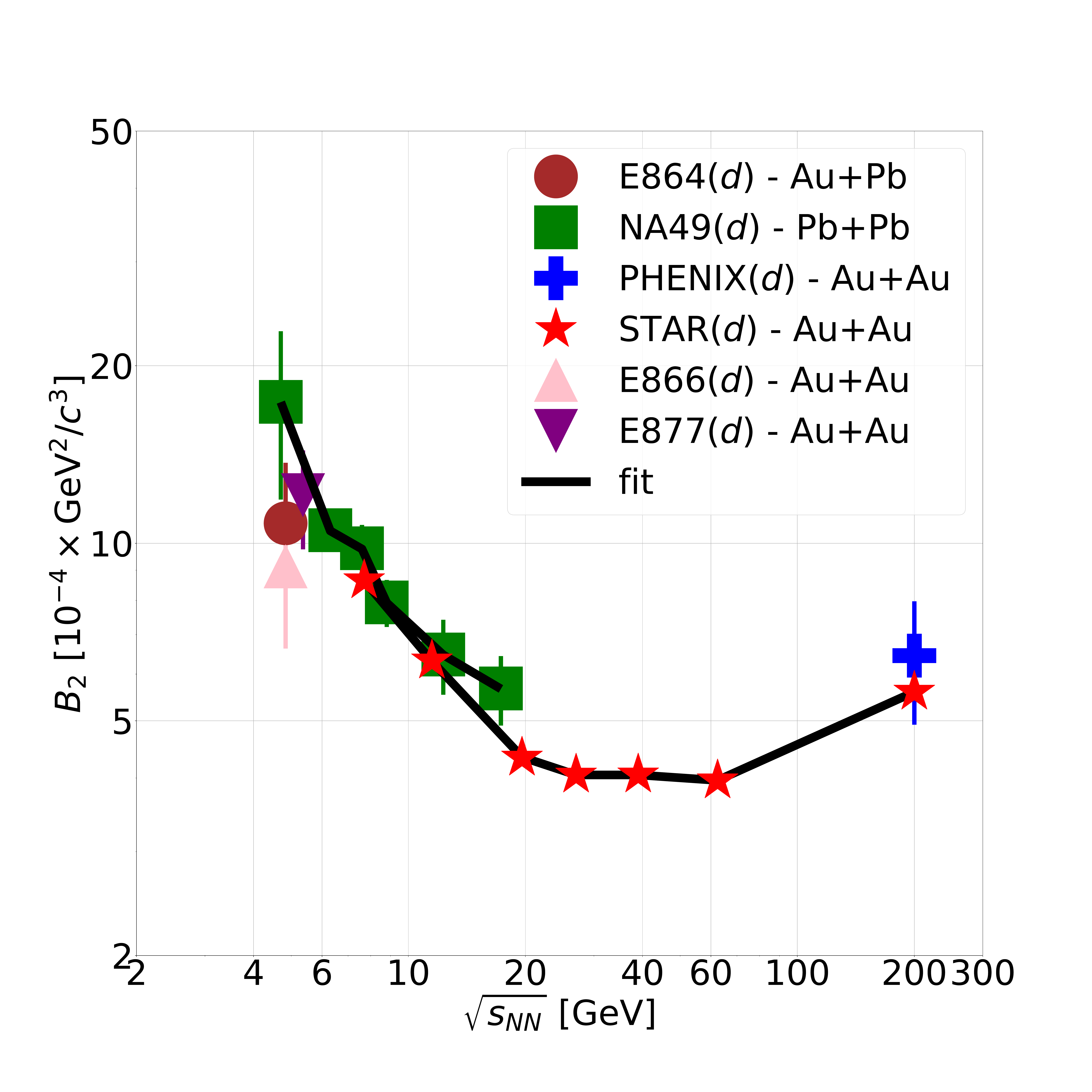}
    \includegraphics[width=\columnwidth]{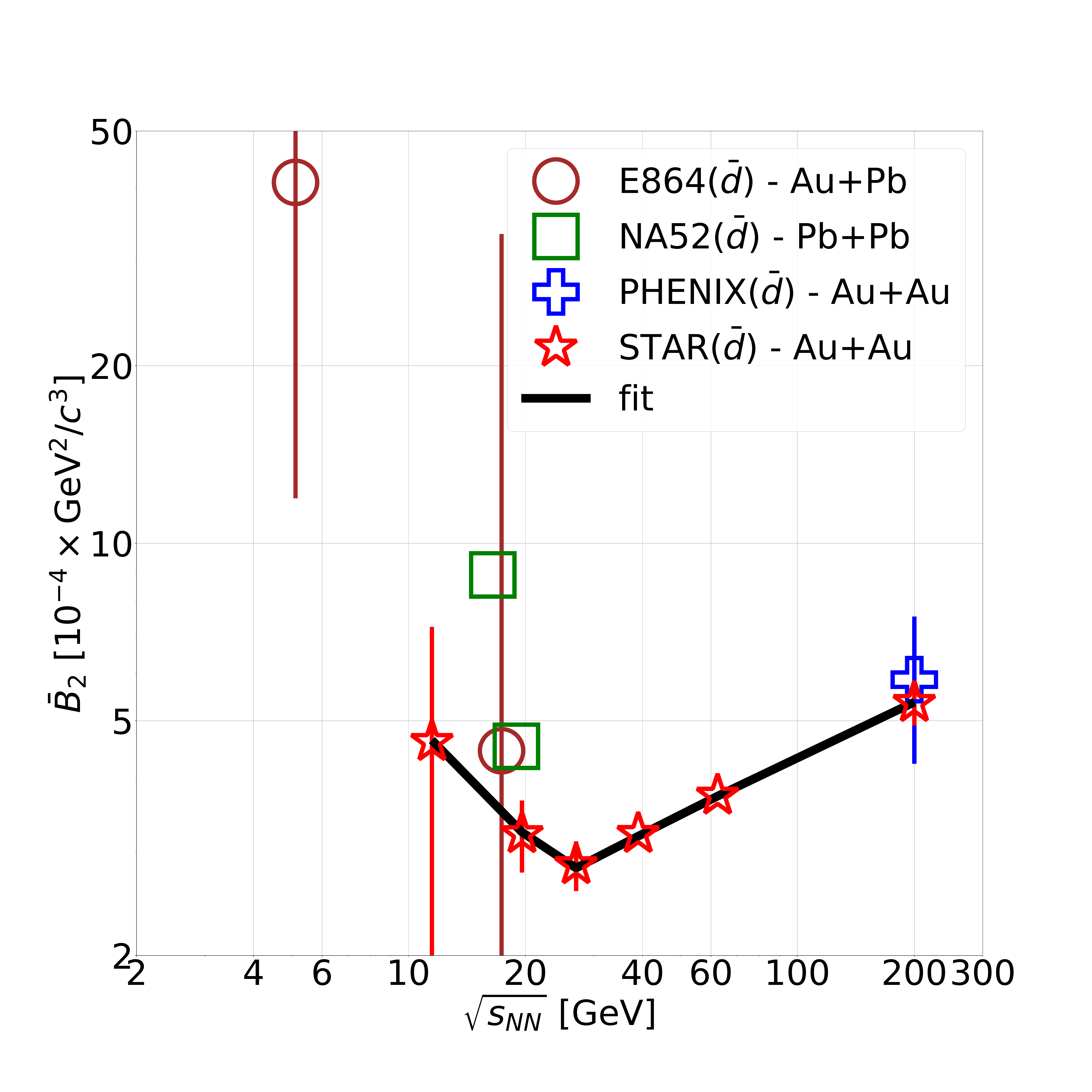}
  \end{subfigure}
\caption{(Color online) The coalescence parameters for deuterons (top) and antideuterons (bottom) extracted from various experimental data  \cite{Ambrosini:1998wg, VanBuren:1999ir, Armstrong:2000gd, Adam:2019wnb} as a function of energy, represented by the symbols. The black lines shows our fit of $B_2$ and $\bar B_2$ using the extracted radii $r_0$ and $r_*$ according to the Mr\'{o}wczy\'{n}ski coalescence model \cite{Mrowczynski:1993cx}.}
\label{fig:experiment}
\end{figure}

To evaluate the source geometries via $r_0$ and $r_*$, we fit the formation rates to the experimental data obtained at different energies. Data on the coalescence parameter $B_2$ \cite{Ambrosini:1998wg, VanBuren:1999ir, Armstrong:2000gd, Adam:2019wnb} is shown in fig.~\ref{fig:experiment} as symbols. We use these to extract the nucleon $r_0$ from the deuteron formation rate $\mathcal{A}$, eq.~\eqref{eq:Ar}. The results of this fitting procedure is shown for the NA49 data and the RHIC-BES data as black lines in fig.~\ref{fig:experiment}. To obtain the values for the antinucleon freeze-out distribution, we assume that the total source size (parametrized by $r_0$ of the nucleons) stays the same and we only need to fit $r_*$ to describe the antideuteron formation.

\begin{figure}[!ht]
\centering
  \begin{subfigure}{}
    \includegraphics[width=\columnwidth]{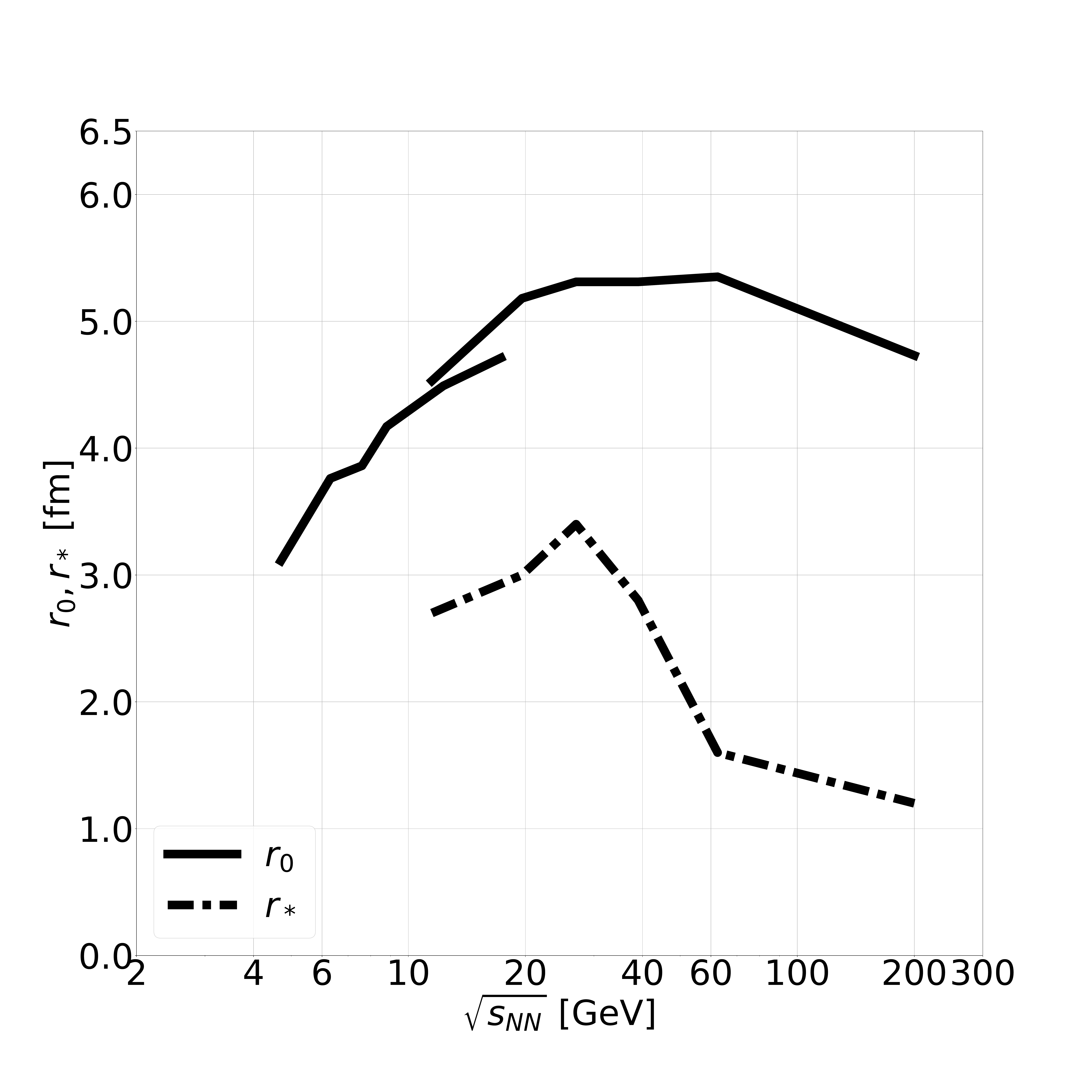}
  \end{subfigure}
\caption{(Color online) The radii $r_0$ and $r_*$ as a function of energy extracted via the coalescence model. The solid line represents $r_0$ for nucleon and antinucleon source. The dashed line represents $r_*$ of the antinucleon source. }
\label{fig:nonsup}
\end{figure}

The results for $r_0$ and $r_*$ are shown in fig.~\ref{fig:nonsup}. At low energies, $\sqrt{s_{NN}} \leq 10$~GeV, the fireball radius $r_0$ grows rapidly with the center-of-mass energy and NA49 data smoothly overlaps with STAR data. A maximum value of $r_0=5.35$~fm is reached around $\sqrt{s_{NN}}=63$~GeV. Toward the higher energy of $200$~GeV, $r_0$ decreases again. This decrease is contrary to the assumption that the fireball radius increases with energy. Since the QGP phase is prominent at this energy, flow effects could significantly affect $B_2$, which is however beyond the scope of our model. The absolute value of the inner radius of the antideuteron production region shows an increase until $\sqrt{s_{NN}}= 27$~GeV, followed by a decrease. This might indicate a nutcracker like shell structure in this energy regime as speculated by Shuryak \cite{Shuryak:1999zz}. Consequently, with increasing beam energy, antinucleons have a higher probability to survive after being produced closer to the center of the fireball. This indicates that towards higher collision energies annihilation becomes less frequent when the system is no longer nucleon-, but pion-dominated.

\section{Validation with UrQMD Data}
\label{sec:urqmd}
Ultrarelativistic Quantum Molecular Dynamics (UrQMD) is a microscopic transport model based on binary scattering of hadrons, resonance excitations, resonance decays, and string dynamics as well as strangeness exchange reactions \cite{Bass:1998ca, Bleicher:1999xi, Graef:2014mra}. We use UrQMD to obtain freeze-out coordinates of nucleons and antinucleons for comparison with our results from the coalescence model. We directly fit the source functions ${\mathcal D}(r)$ and $\bar{\mathcal D}(r)$ to transverse radius ($r_T$) distributions of (anti-)nucleons from central events, $1/r_T~dN/dr_T$. We divide by $r_T$ to account for the approximately cylindrical geometry at mid-rapidity in the center-of-mass frame. In this way, we are able to extract $r_0$ and $r_*$ directly from the simulation and independent of the coalescence parameter $B_2$. 

\begin{figure}[!ht]
\centering
\includegraphics[width=\columnwidth]{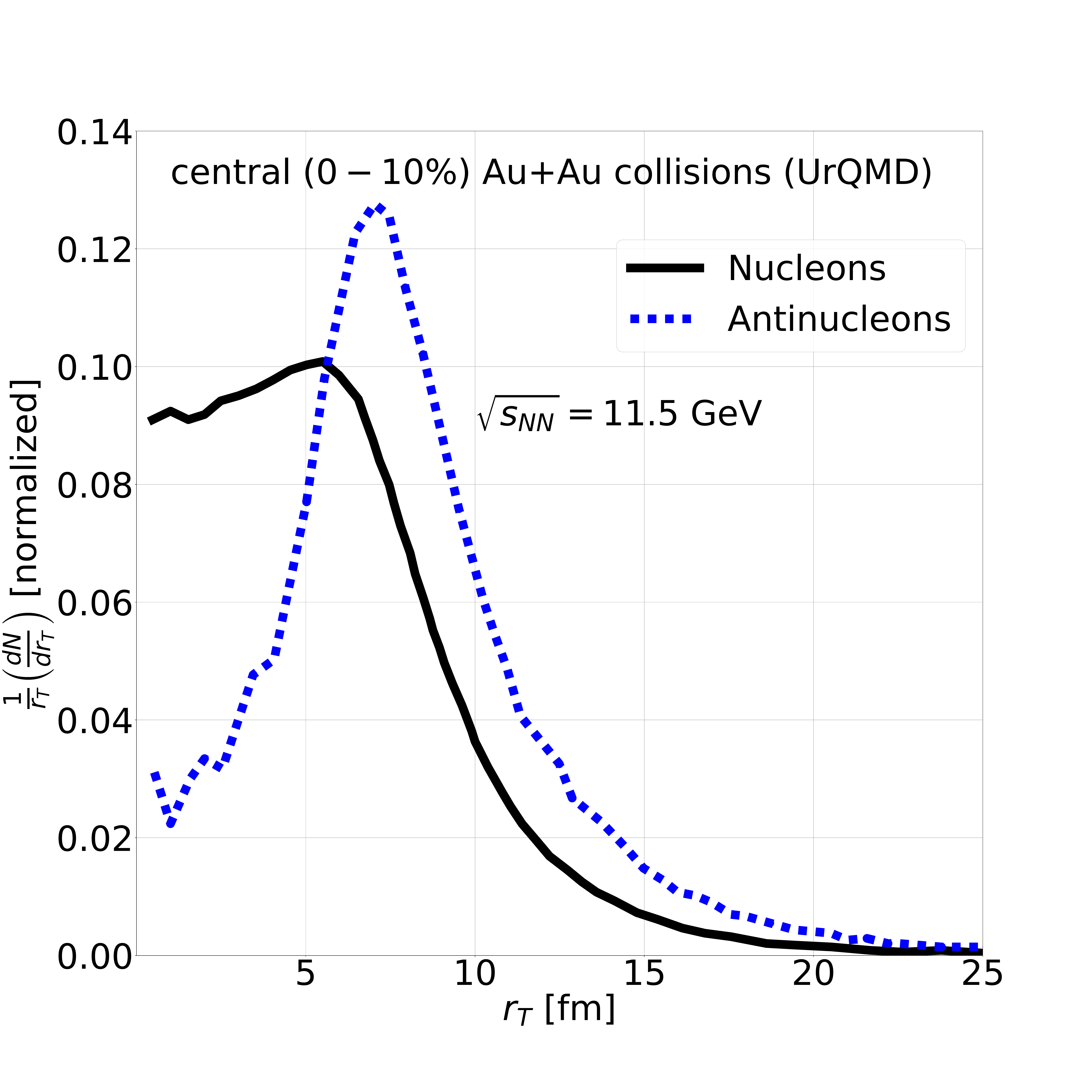}
\includegraphics[width=\columnwidth]{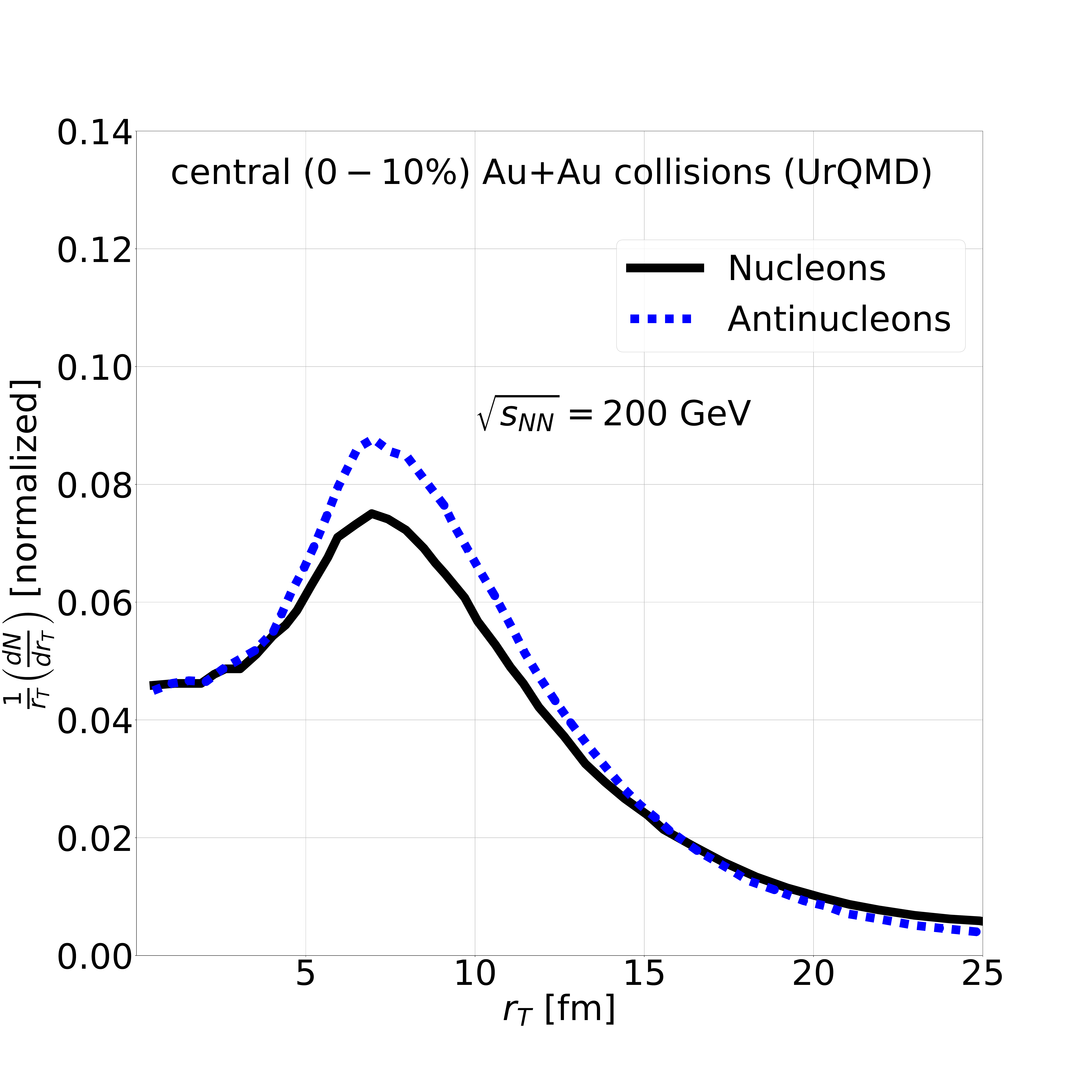}
\caption{(Color online) The (anti-)nucleon distributions along the transverse radius $r_T$ using UrQMD in central $(0-10\%)$ Au+Au collisions at $\sqrt{s_{NN}} = 11.5$~GeV (top) and $ 200$~GeV (bottom). The curves are normalized such that the integral over $r_T$ yields unity.}
\label{fig:distribution}
\end{figure}

We analyze central ($0-10$\%) Au+Au collisions at the STAR center-of-mass energies, $\sqrt{s_{NN}}=$ 7.7, 11.5, 14.5, 19.6, 27, 39, 62.4, and 200~GeV.
Figure \ref{fig:distribution} shows the $r_T$-distribution of (anti-)nucleons at $\sqrt{s_{NN}}=11.5$ and $200$~GeV. The curves are normalized to unity. As expected from our analysis above, we notice a qualitatively different behavior at the two energies. At $11.5$~GeV, the number of antinucleons is strongly depleted near the center of the fireball while the number of nucleons is approximately constant over the transverse area. This clearly reflects the effect of nucleon-antinucleon annihilation in the baryon-rich environment. The assumptions of a Gaussian shape and a Gaussian suppression region apply quite well. The UrQMD distributions thus support the central idea behind Mr\'{o}wczy\'{n}ski's coalescence model \cite{Mrowczynski:1993cx}.
At $200$~GeV, the distributions of nucleons and antinucleons resemble each other, indicating a similar geometry of the nucleon and antinucleon sources. Both curves decrease towards small $r_T$ demonstrating lower (anti-)nucleon abundance in the central fireball region. Such dilution is compatible with the notion of pion domination at high energies.

\begin{figure}[!t]
\centering
\includegraphics[width=\columnwidth]{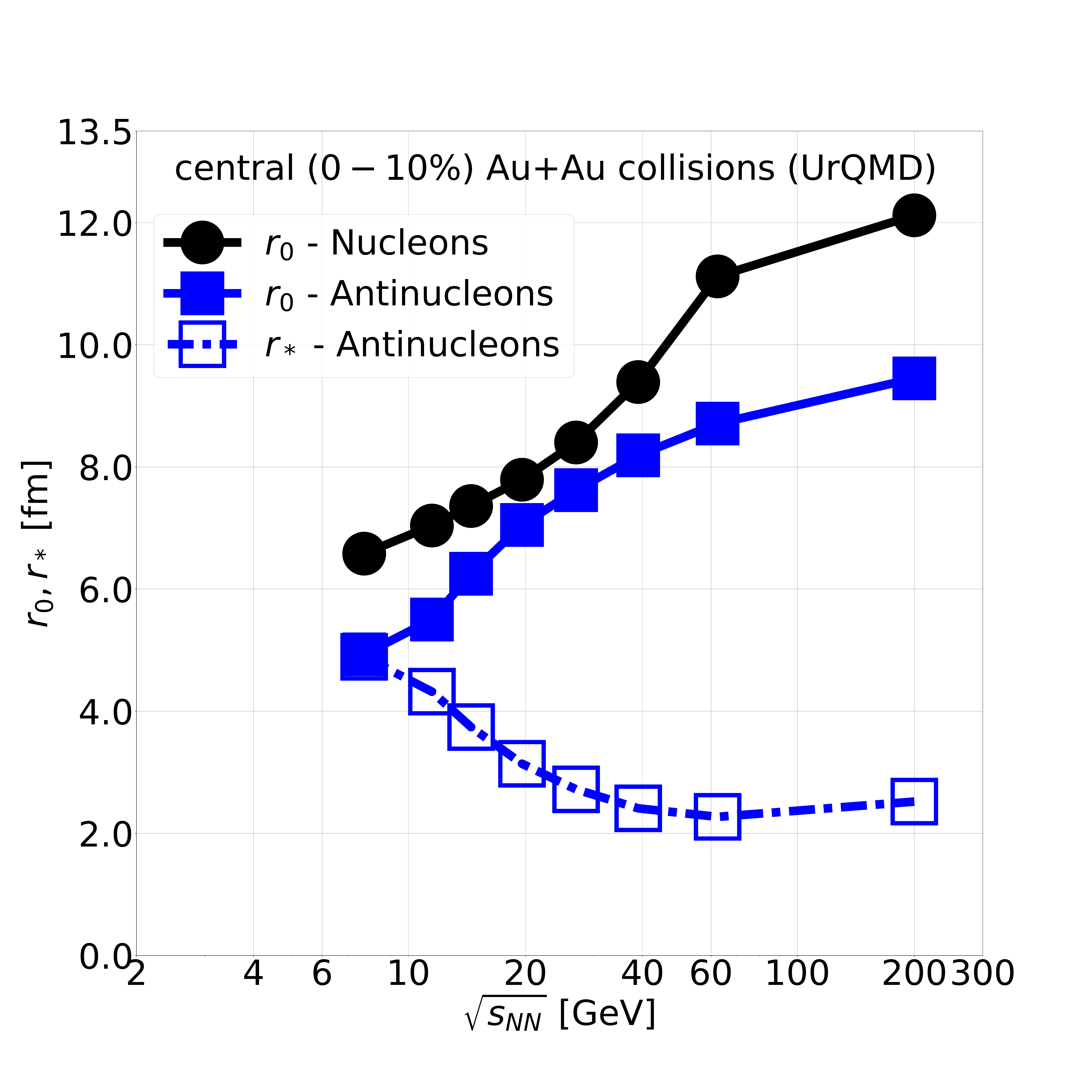}
\caption{(Color online) $r_0$ and $r_*$ of (anti-)nucleon source as a function of energy extracted from UrQMD. The circle symbol represents the nucleon, the square symbol the antinucleon source.}
\label{fig:urqmd-r0rs}
\end{figure}

The extracted values for $r_0$ and $r_*$ for the (anti-)nucleon source functions are shown in fig.~\ref{fig:urqmd-r0rs}. We find that the resulting $r_0$ values for nucleons and antinucleons are close together with a very similar energy dependence. This finding corroborates the assumption that we used in the coalescence model analysis. In contrast to the coalescence model values extracted from the data, there is no peak for $r_*$ at 27~GeV and no slight decrease in $r_0$ at 200~GeV.  Generally, the UrQMD results are larger by a factor of around 2. Such systematic deviations are to be expected because we compare the Mr\'{o}wczy\'{n}ski model, assuming instantaneous (anti-)nucleon freeze-out, to the dynamical model UrQMD with a time dependent freeze-out. 

\begin{figure}[!t]
\centering
\includegraphics[width=\columnwidth]{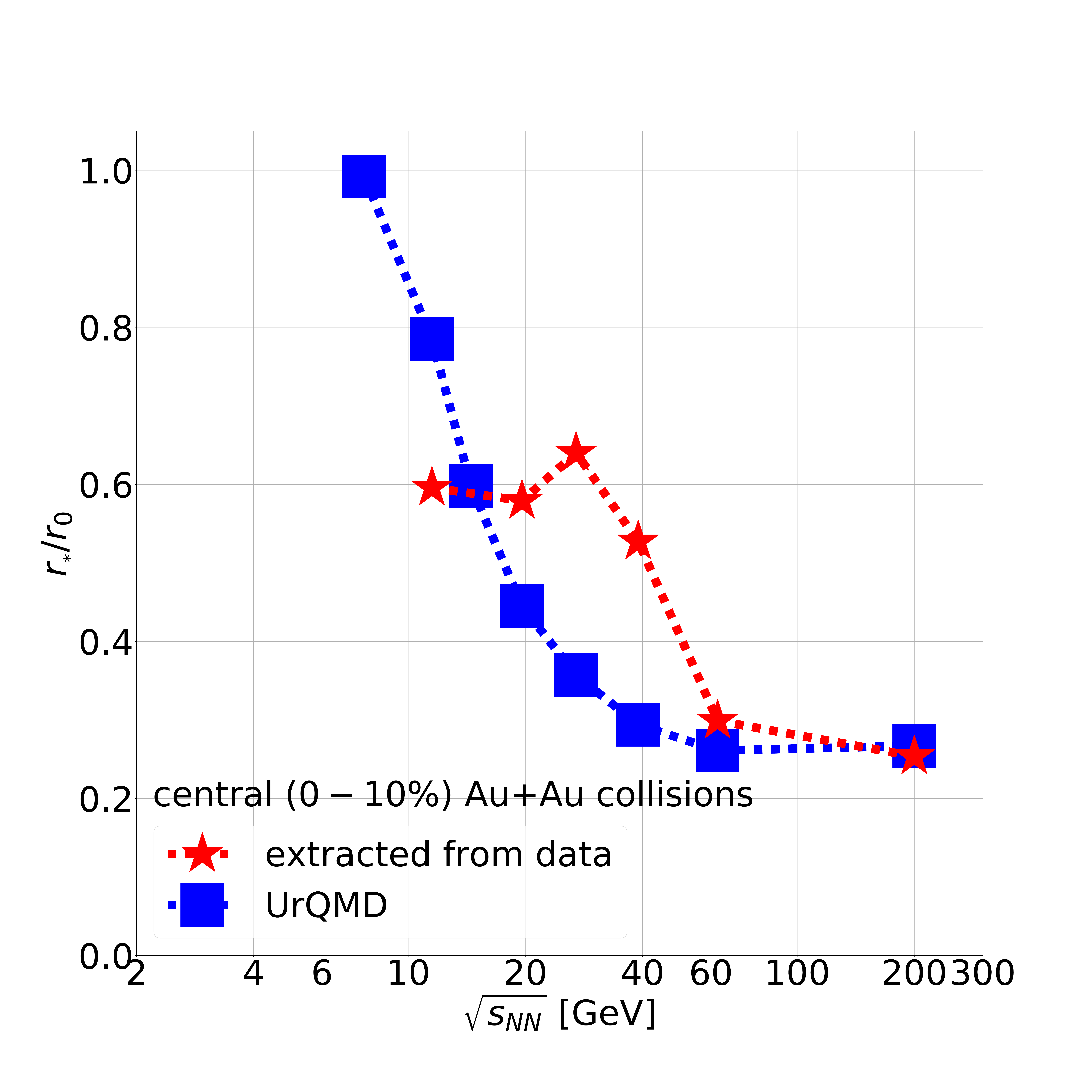}
\caption{(Color online) Comparison of energy dependence of antinucleon source ratio $r_*/r_0$ between UrQMD (square symbol) and coalescence model (star symbol).}
\label{fig:urqmd-ratio}
\end{figure}

Nevertheless, as we can see from fig.~\ref{fig:urqmd-ratio}, the extracted $r_*/r_0$ ratios from the Mr\'{o}wczy\'{n}ski coalescence model and from UrQMD show a similar trend to decrease with higher beam energy. This indicates that at low energies, antinucleons are solely emitted close to the surface of the fireball and that the relative region of suppression shrinks with increasing energy. Note that for $\sqrt{s_{NN}} \leq 5$~GeV, no reliable data for the antideuteron coalescence parameter is available. We estimate that here $r_0$ and $r_*$ should be very close to each other resulting in a ratio $r_*/r_0 \simeq 1$, in agreement with the UrQMD result.

This investigation justifies the main assumption of the coalescence model that most antinucleons are emitted on the outer shell of the fireball, while nucleons are emitted from the whole source volume.

\section{Conclusions}
\label{sec:concl}
We studied the energy dependence of the formation geometry of deuterons and antideuterons in central heavy-ion collisions.
According to Mr\'{o}wczy\'{n}ski's coalescence model, nucleons are emitted in the whole fireball of source radius $r_0$, while antinucleons are suppressed in the center, characterized by the suppression radius $r_*$. Assuming Gaussian distributions, we extracted $r_0$ from the deuteron coalescence parameter $B_2$. Subsequently, we obtained $r_*$ from the antideuteron $\bar{B}_2$. 

We find that the relative suppression of antinucleons recedes with increasing energy. In other words, antideuterons can form closer to the fireball center as well. At the highest energies around $200$~GeV, the formation geometries of deuterons and antideuterons agree. We argue that the reason is the transition from a nucleon- to pion-dominated fireball when the center-of-mass energy increases.
This interpretation is backed up by an UrQMD analysis of the transverse radius profiles of nucleons and antinucleons at kinetic freeze-out.

Let us finally point to an interesting non-monotonous structure in the $r_*/r_0$ ratio, see fig.~\ref{fig:urqmd-ratio} (bottom). The apparent local maximum of $r_*/r_0$ extracted from the data is around $\sqrt{s_{NN}}\simeq 27$~GeV. This could be a sign of a change in the equation of state close to a QCD critical point and is also reflected in the coalescence parameters measured by STAR \cite{Adam:2019wnb}. It would be interesting to further test this idea when more statistics on antideuteron formation at low energies are available.

\section*{Acknowledgements}
This work was supported by the Development and Promotion of Science and Technology Talents Project (DPST) - Royal Thai  Government Scholarship, Suranaree University of Technology (SUT), the Deutscher Akademischer Austausch Dienst (DAAD), the Stiftung Polytechnische Gesellschaft Frankfurt am Main, the Helmholtz International Center for FAIR (HIC for FAIR) within the LOEWE program launched by the State of Hesse, and the COST Action CA15213 (THOR). The computational resources were provided by the Center for Scientific Computing (CSC) at the Goethe-Universit\"at Frankfurt. We would also like to express our gratitude to our colleagues in Frankfurt and at SUT for their hospitality, enthusiastic engagement in the discussions and valuable suggestions.

\bibliographystyle{epj}
\bibliography{bibliography}

\end{document}